\documentclass{elsart}
\usepackage{epsfig}

\def\DAF{DA\char8NE}  \def\epm{\ifm{e^+e^-}}
\def\ifm#1{\relax\ifmmode#1\else$#1$\fi}
\def\pic{\ifm{\pi^+\pi^-}}
\def\toP{\ifm{\rightarrow}}
\def\ff{$\phi$--factory}  \def\f{\ifm{\phi}}
\def\K{\ifm{K}}
\def\pienu{\ifm{\pi^\pm e^\mp\bar\nu(\nu)}}
\def\ab{\ifm{\sim}}  \def\x{\ifm{\times}}
\def\pt#1,#2,{\ifm{#1\x10^{#2}}}
\makeatletter
\newdimen\z@ \z@=0pt 
\newskip\z@skip \z@skip=0pt plus0pt minus0pt
\def\m@th{\mathsurround=\z@}
\def\ialign{\everycr{}\tabskip\z@skip\halign} 
\def\eqalign#1{\null\,\vcenter{\openup\jot\m@th
  \ialign{\strut\hfil$\displaystyle{##}$&$\displaystyle{{}##}$\hfil
    \crcr#1\crcr}}\,}
\makeatother
\newcommand{\aff}[2]{Dipartimento di Fisica dell'Universit\`a #1 e Sezione INFN, #2, Italy.}
\newcommand{\affd}[1]{Dipartimento di Fisica dell'Universit\`a e Sezione INFN, #1, Italy.}
\newcommand{\kl}{\mbox{$K_L$}}
\newcommand{\ks}{\mbox{$K_S$}}
\newcommand{\Pe}{\ensuremath{e}}
\newcommand{\Pem}{\ensuremath{e^-}}
\newcommand{\Pep}{\ensuremath{e^+}}
\newcommand{\Pemp}{\ensuremath{e^\mp}}
\newcommand{\Pmump}{\ensuremath{\mu^\mp}}
\newcommand{\Pnu}{\ensuremath{\nu}}
\newcommand{\Pnubar}{\ensuremath{\bar{\nu}}}
\newcommand{\Pphi}{\ensuremath{\phi}}
\newcommand{\Ppi}{\ensuremath{\pi}}
\newcommand{\Ppim}{\ensuremath{\pi^-}}
\newcommand{\Ppip}{\ensuremath{\pi^+}}

\newcommand{\tzero}{\ensuremath{T_{0}}}

\newcommand{\eV}{{e\kern-.07em V}}
\newcommand{\MeV}{{\rm \,M\eV}}
\newcommand{\GeV}{{\rm G\eV}}
\newcommand{\ps}{{\rm \,ps}}
\newcommand{\ns}{{\rm \,ns}}
\newcommand{\mm}{{\rm \,mm}}
\newcommand{\cm}{{\rm \,cm}}
\newcommand{\m}{{\rm \,m}}
\newcommand{\um}{\ensuremath{\mathrm{\mu m}}}
\newcommand{\T}{{\rm \,T}}
\newcommand{\Lpb}{\ensuremath{\rm pb^{-1}}}
\newcommand{\dsdq}{\ensuremath{\Delta S\!=\!\Delta Q}}
\newcommand{\dsnotdq}{\ifm{\Delta S\!=\!-\Delta Q}}
\newcommand{\phikskl}{\ensuremath{\phi\rightarrow K_S K_L}}
\newcommand{\epemphikskl}{\ensuremath{e^+e^-\rightarrow\phi\rightarrow K_S K_L}}
\newcommand{\DKSeIII}{\ensuremath{K_S\rightarrow\pi e \nu}}
\newcommand{\DKSpippim}{\ensuremath{K_S\rightarrow\pi^+\pi^-}}

\newcommand{\DKLeIII}{\ensuremath{K_L\rightarrow\pi e \nu}}
\newcommand{\DKLpippimpin}{\ensuremath{K_L\rightarrow\pi^+\pi^-\pi^0}}

\newcommand{\Dphipippimpin}{\ensuremath{\phi\rightarrow\pi^+\pi^-\pi^0}}
\newcommand{\BR}[1]{\ensuremath{\mathrm{BR}(#1)}}
\newcommand{\Emiss}{\ensuremath{E_\mathrm{miss}}}
\newcommand{\Pmiss}{\ensuremath{p_\mathrm{miss}}}
\newcommand{\SN}[2]{\ensuremath{#1\times10^{#2}}}
\newcommand{\VS}[2]{\ensuremath{#1\pm#2}}
\newcommand{\Fig}[1]{Fig.~#1}
\begin{document}
\begin{frontmatter}
\title{Measurement of the branching fraction for the decay $\DKSeIII$}
\collab{The KLOE Collaboration}
\author[Na]{A.~Aloisio},
\author[Na]{F.~Ambrosino},
\author[Frascati,Moscow]{A.~Andryakov},
\author[Frascati]{A.~Antonelli},
\author[Frascati]{M.~Antonelli},
\author[Roma3]{C.~Bacci},
\author[Frascati]{G.~Bencivenni},
\author[Frascati]{S.~Bertolucci},
\author[Roma1]{C.~Bini},
\author[Frascati]{C.~Bloise},
% ...
\address[Roma2]{\aff{``Tor Vergata''}{Roma}}
\address[Na]{Dipartimento di Scienze Fisiche dell'Universit\`a ``Federico II'' e Sezione INFN,
Napoli, Italy}
\address[Moscow]{Permanent address: Institute for Theoretical and Experimental Physics, Moscow, Russia.}
\address[Frascati]{Laboratori Nazionali di Frascati dell'INFN, Frascati, Italy.}
\address[Roma3]{\aff{``Roma Tre''}{Roma}}
\address[Trieste]{\affd{Trieste}}
\address[Roma1]{\aff{``La Sapienza''}{Roma}}
\address[Lecce]{\affd{Lecce}}
\address[Pisa]{\affd{Pisa}}
\address[Bari]{\affd{Bari}}
\address[Beijing]{Permanent address: Institute of High Energy Physics, CRS, Beijing, China.}
\address[StonyBrook]{Physics Department, State University of New York at Stony Brook, USA.}
\address[Karlsruhe]{Institut f\"ur Experimentelle Kernphysik, Universit\"at Karlsruhe, Germany.}
\address[Energ]{Dipartimento di Energetica dell'Universit\`a ``La Sapienza'', Roma, Italy.}
\address[Virginia]{Physics Department, University of Virginia, USA.}
\corauth[cor1]{Corresponding author: Claudio Gatti
INFN - LNF, Casella postale 13, 00044 Frascati (Roma), 
Italy; tel. +39-06-94032727, e-mail claudio.gatti@lnf.infn.it}
\corauth[cor2]{Corresponding author: Tommaso Spadaro
INFN - LNF, Casella postale 13, 00044 Frascati (Roma), 
Italy; tel. +39-06-94032696, e-mail tommaso.spadaro@lnf.infn.it}
%
\begin{abstract}
We present a measurement of the branching ratio \BR{\K\toP\pienu} performed using the KLOE detector. 
\ks-mesons are produced in the reaction \epemphikskl\ at the \DAF\ collider. In a sample of 
$\ab\!\pt 5,6,$ \ks-tagged events 
we find \VS{624}{30} semileptonic \ks\ decays. Normalizing to the \ks\toP\pic\ count in the same 
data sample, we obtain $\BR{\DKSeIII}\!=\!\pt(6.91\pm0.37),-4,$, in agreement with the Standard Model 
expectation.
\end{abstract}
\end{frontmatter}
In the Standard Model, there are only \dsdq\ transitions at tree level. \dsnotdq\ transitions exist at higher
order, but are suppressed by a factor of $\ab\!10^{-6}$--$10^{-7}$~\cite{Dib1991ps,Luke1991dh}. Under this
assumption, together with $TCP$-invariance~\cite{maia,paf}, it follows that
$\Gamma(\DKSeIII)\!=\!\Gamma(\DKLeIII)$ to very high accuracy. Using the values of $\BR{\DKLeIII}$ and
$\tau_{S}/\tau_{L}$ from ref~\cite{Groom:2000in}, one obtains $\BR{\DKSeIII}\!=\!\SN{({\VS{6.70}{0.07}})}{-4}$.
Precise measurements of the \ks\ semileptonic decay rate are important for checking the above assumptions. 
Until recently, no experimental information was available about the semileptonic decay of the \ks-meson. 
This is in large part due to the smallness of the branching ratio and to the difficulty of isolating \pienu\ 
decays from \pic\ decays. Both decays are observed as two charged tracks,
but the latter are $\ab\!1000$ times more abundant. Good
particle identification is necessary for isolating a pure sample of \pienu\ events. The CMD-2 
collaboration at VEPP-2M operating at the \f\ peak has observed $75\pm13$ semileptonic 
\ks\ decays~\cite{CMD2kse3}, with limited separation of signal and background. 
They find $\BR{\DKSeIII}\!=\!\SN{({\VS{7.2}{1.4}})}{-4}$. With KLOE~\cite{Kloep,Kloet}, we have considerably 
improved on this result.

Our measurement is performed with kaons from \phikskl\ decays. KLOE operates at \DAF~\cite{Dafneref}, an
\epm\ collider also known as the Frascati \ff. \f-mesons are produced in small angle (25 mrad) collisions of
equal energy electrons and positrons, giving the \f\ a small transverse momentum component
in the horizontal plane, $p_\Pphi\!\ab\!13\MeV/c.$
The main advantage of studying kaons at a \ff\ is that \f-mesons decay
$\ab\!34\%$ of the time into neutral kaons. \kl's and \ks's are produced almost
back-to-back in the laboratory, with mean decay paths $\lambda_{\rm L}\!\ab\!340$\cm\ and
$\lambda_{\rm S}\!\ab\!0.6$\cm,
respectively. Detection of a long-lived kaon therefore tags the production of a \ks-meson and gives its
direction and momentum. 
The contamination from $\kl\kl\gamma$ and $\ks\ks\gamma$ final states
is negligible for our measurement~\cite{Dunietz:1987jf,brownclose}.
Since the branching ratio for \DKSpippim\ is known with an accuracy of $\ab0.4\%$~\cite{Groom:2000in}, the
\DKSeIII\ branching ratio is evaluated by normalizing the number of signal events to the number of \DKSpippim\
events in the same data set. This allows cancellation of the uncertainties  coming from the integrated
luminosity, the \f\ production cross section, and the tagging efficiency.
Both charge states $\Ppip\Pem\Pnubar$ and
$\Ppim\Pep\Pnu$ are included in the measured branching ratio. The measurement is based on an integrated
luminosity of
17\,\Lpb\ at the \f\ peak collected during the year 2000, corresponding to $\ab\!\pt5,7,$ produced \f-mesons.

\begin{figure}[h]
  \center
    \epsfig{file=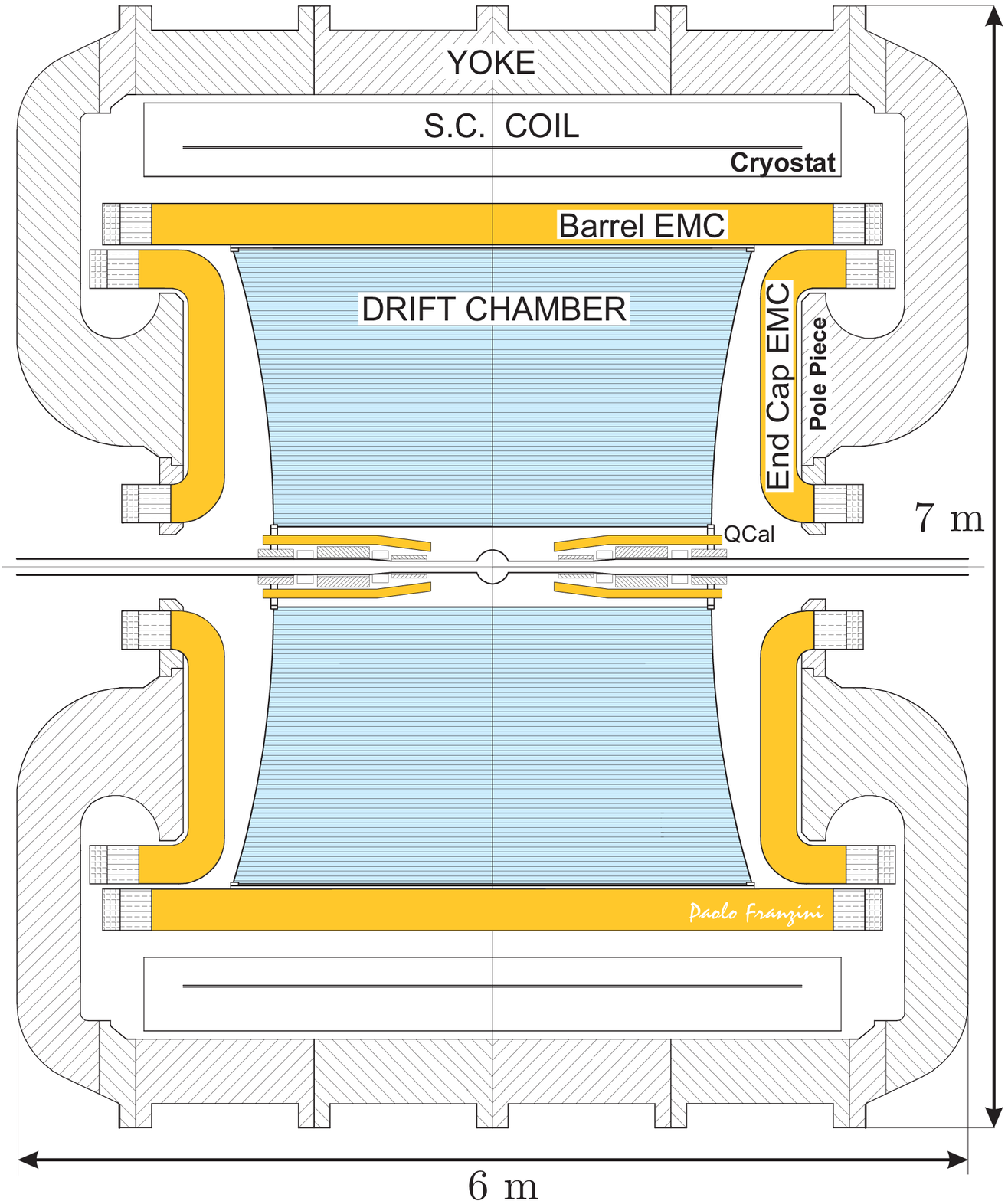,width=0.35\textwidth}
\caption{Vertical cross section of the KLOE detector.}
\label{det}
\end{figure}

The KLOE detector (\Fig{\ref{det}}) 
consists of a large cylindrical drift chamber surrounded by a lead-scintillating fiber
sampling calorimeter. A superconducting coil outside the calorimeter 
provides a 0.52\T\ field. The drift chamber~\cite{DCnim}, 
4\m\ in diameter and 3.3\m\ long, has 12\,582 all-stereo sense wires and 37\,746 aluminium field wires. 
The chamber shell is 
made of carbon fiber-epoxy composite and the gas used is a 90\% helium, 10\% isobutane mixture. These 
features maximize transparency to photons and reduce \kl\toP\ks\ regeneration. The position resolutions are
$\sigma_{xy}\!\sim\!150\,\um$ and $\sigma_z\!\sim\!2\mm.$ The momentum resolution is 
$\sigma(p_{\perp})/p_{\perp}\!\leq\!0.4\%$. Vertices are reconstructed with a spatial resolution of $\sim\!3\mm$. 
The calorimeter~\cite{EmCnim}, divided into a barrel and two endcaps, covers 98\% of the 
solid angle. The energy resolution is $\sigma_E/E\!=\!5.7\%/\sqrt{E (\GeV)}$ and the timing resolution 
is $\sigma_t\!=\!54\ps/\sqrt{E (\GeV)}\oplus50\ps.$ The trigger~\cite{TRGprop} uses calorimeter and chamber 
information. For the present analysis, the trigger relies entirely on 
calorimeter information. Two local energy deposits above threshold 
($50$\MeV\ on the barrel, $150$\MeV\ on the endcaps) are required.
The trigger time has a large spread with respect to the bunch crossing time. It is however 
synchronized with the machine RF divided by 4, $T_{\rm sync}\!=\!10.8\ns$, with an accuracy of 50\ps. The time
\tzero\ of the bunch crossing producing an event is determined after event reconstruction. 

About half of the \kl-mesons reach the calorimeter, where most interact. A \kl\ interaction is called 
a \kl-{\it crash} in the
following. A \kl-crash is identified as a local energy deposit with energy above 100\MeV\ and a time
of flight indicating low velocity: $\beta\!\ab\!0.218$. The coordinates of the energy deposit determine the \kl's
direction to $\ab\!20$\,mrad and its momentum $\mathbf{p}_L$, 
which is weakly dependent on the \kl\ direction because of the motion
of the  \f-meson. A \kl-crash thus tags the production of a \ks\ of momentum
$\mathbf{p}_S\!=\!\mathbf{p}_\Pphi-\mathbf{p}_L$. \ks-mesons are tagged with an overall efficiency of $\ab\!30\%$.
Both \DKSeIII\ and \DKSpippim\ decays are selected from this tagged sample. Event selection consists of fiducial cuts,
particle identification by time of flight, and kinematic closure.

Identification of \DKSpippim\ decays requires two tracks of opposite curvature. Tracks must extrapolate to the interaction
point (IP) within KLOE resolutions. The reconstructed momenta and polar angles must lie in the intervals 
$120\MeV/c\!<\!p\!<\!300\MeV/c$ and $30^{\circ}\!<\!\theta\!<\!150^{\circ}$. A cut in $(p_{\perp},p_{\parallel})$ selects 
non-spiralling tracks. 1\,636\,604 \DKSpippim\ events have been found.
Contamination due to \ks\ decays other than \DKSpippim\ is well below the per-mil level and is ignored. 

Identification of \DKSeIII\ events also begins with two tracks of opposite curvature. Tracks must extrapolate
{\it and form a vertex} at the IP. 
The invariant mass $M_{\mathrm{\pi\pi}}$ of the pair calculated assuming both tracks are pions must be smaller 
than 490\MeV. This rejects $\ab\!95\%$ of the \pic\ decays.

\begin{figure}[h]
  \center
    \epsfig{file=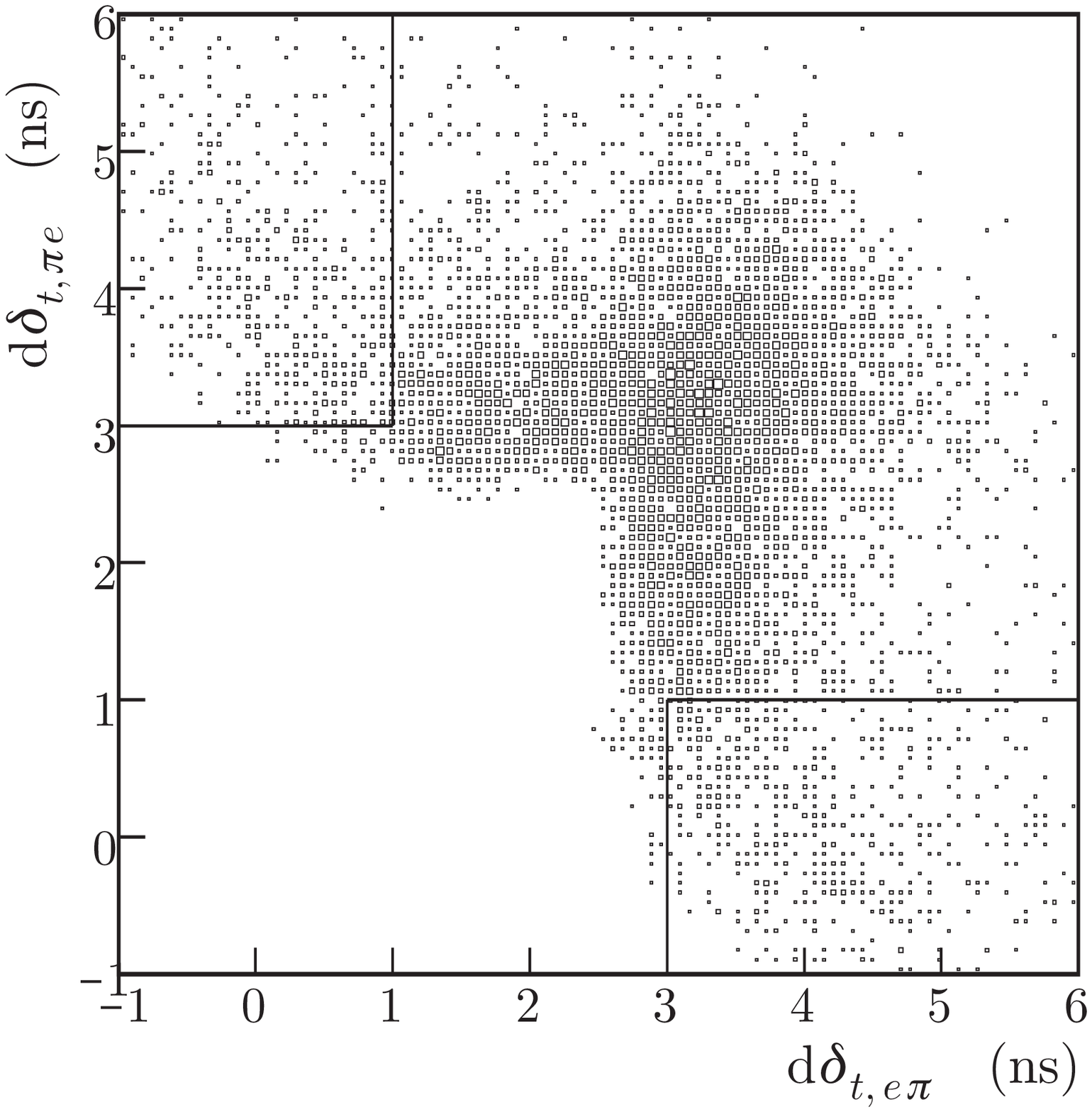, width=0.35\textwidth}
    \epsfig{file=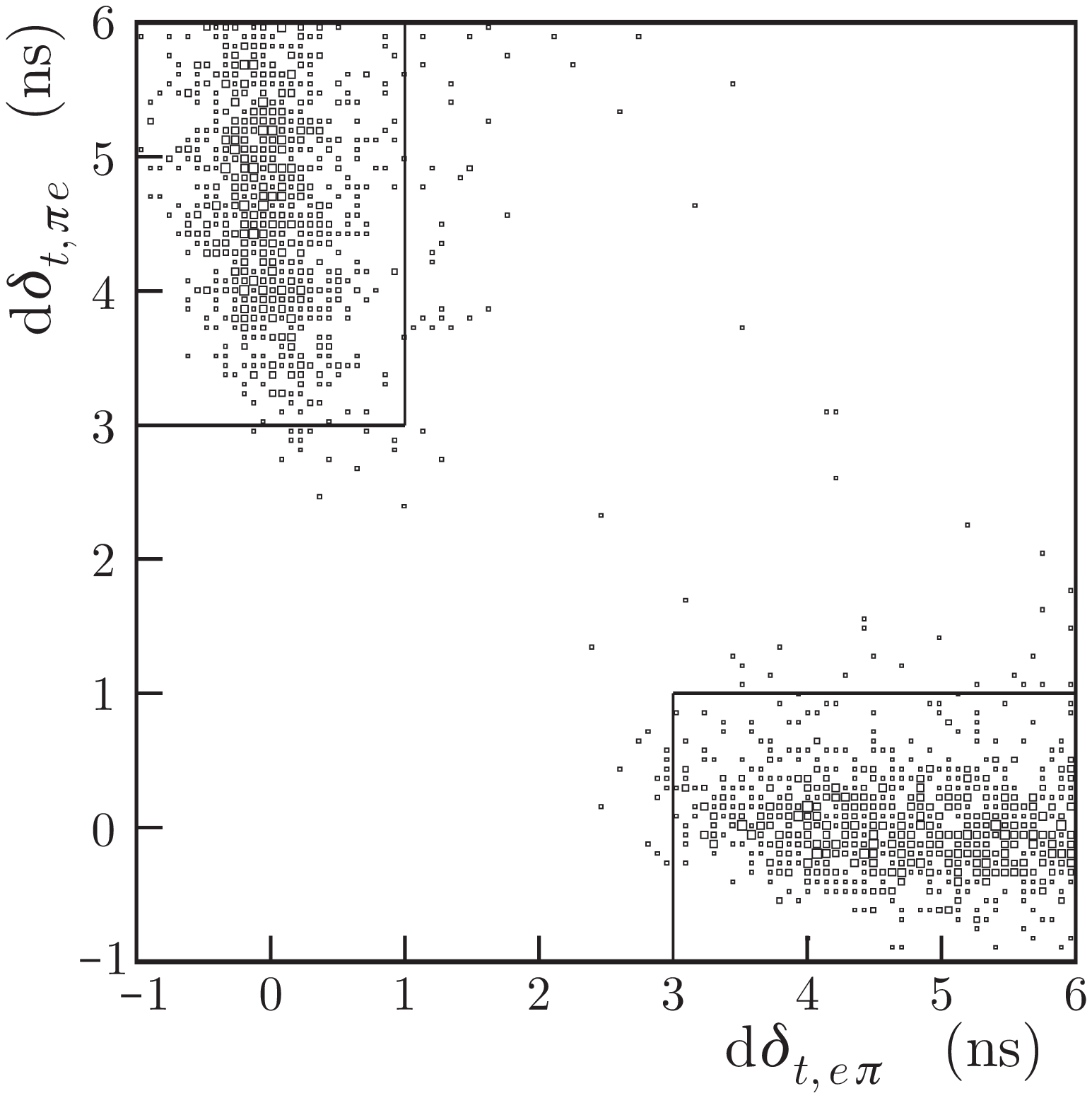, width=0.35\textwidth}
\caption{Scatter plot of the time differences $d\delta_{t\mathrm{,\Ppi\Pe}}$ vs $d\delta_{t\mathrm{,\Pe\Ppi}}$
for $\Pe\Ppi$ and $\Ppi\Pe$ mass assignments for \DKSpippim\ (left) and \DKSeIII\ (right) Monte Carlo events.}
\label{cut}
\end{figure}

Electrons and pions are discriminated by time of flight (TOF). Tracks are therefore required to be
associated with calorimeter energy clusters. For each track,
we compute the difference $\delta_t(m)\!=\!t_{\rm cl}-L/c\beta(m)$
using the cluster time $t_{\rm cl}$ and the track length $L$. The velocity
$\beta$ is computed from the track momentum for each mass hypothesis, $m\!=\!m_{e}$ and $m\!=\!m_{\Ppi}$.
In order to avoid uncertainties due to the determination of \tzero,
we make cuts on the two-track difference
$$
d\delta_{t\mathrm{,ab}}=\delta_t(m_{a})_{1}-\delta_t(m_{b})_{2}\mbox{,}
$$
where the mass hypothesis $m_{a(b)}$ is used for the track 1(2). This difference is zero for the correct
mass assignments.
An additional fraction of \DKSpippim\ events is rejected by requiring
$|d\delta_{t\mathrm{,\Ppi\Ppi}}|>1.5$\ns. The differences
$d\delta_{t\mathrm{,\Ppi\Pe}}$ and $d\delta_{t\mathrm{,\Pe\Ppi}}$
are calculated for events surviving the previous cut. The scatter plot
of the two variables is shown in \Fig{\ref{cut}} for Monte Carlo events. The cuts applied on
these time differences for the selection of \DKSeIII\ events are illustrated in the figure:
$|d\delta_{t\mathrm{,\Ppi\Pe}}|<1$\ns, $d\delta_{t\mathrm{,\Pe\Ppi}}>3$\ns; or
$|d\delta_{t\mathrm{,\Pe\Ppi}}|<1$\ns, $d\delta_{t\mathrm{,\Ppi\Pe}}>3$\ns.

\begin{figure}
\center
\epsfig{file=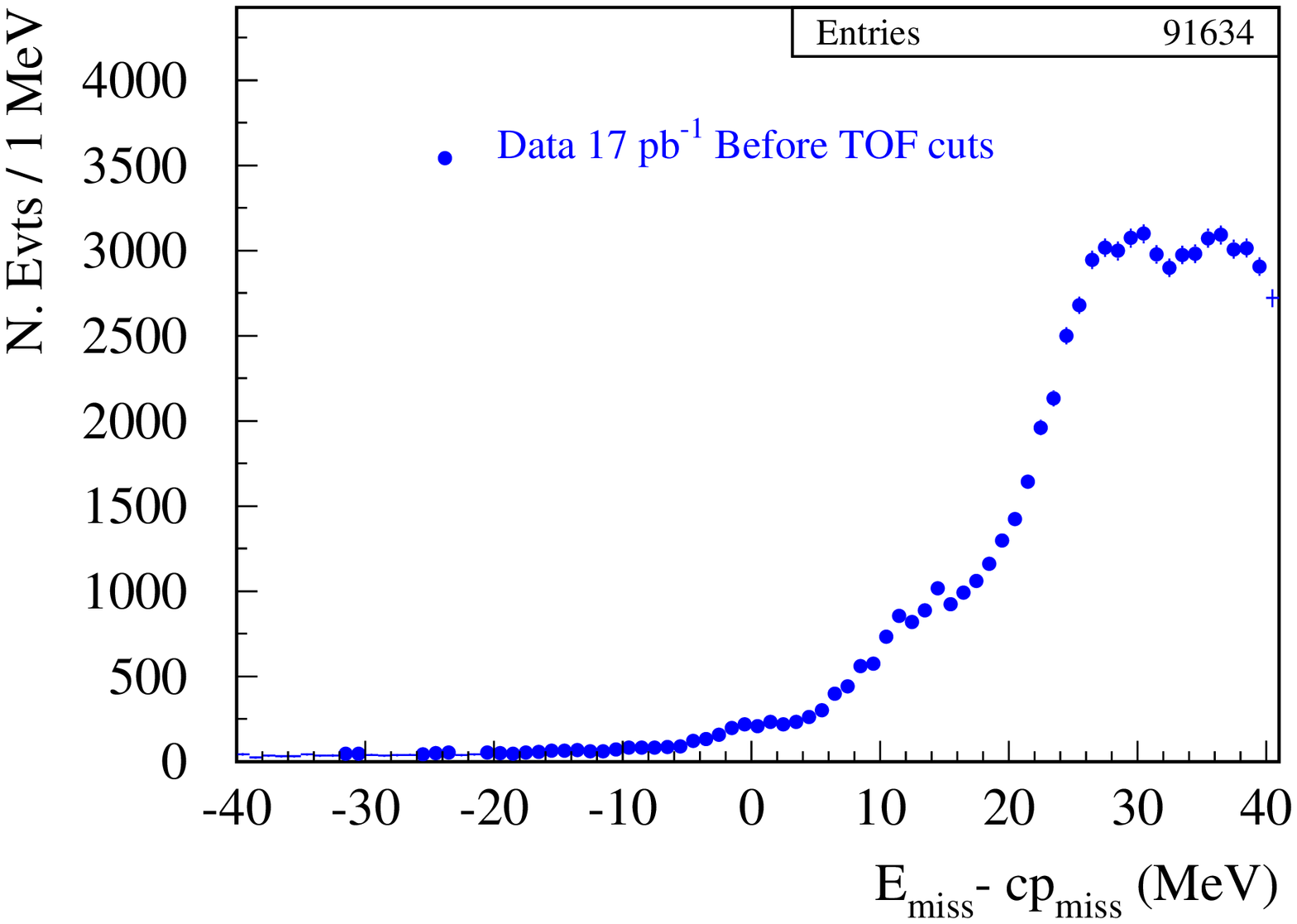, width=0.4\textwidth}
\epsfig{file=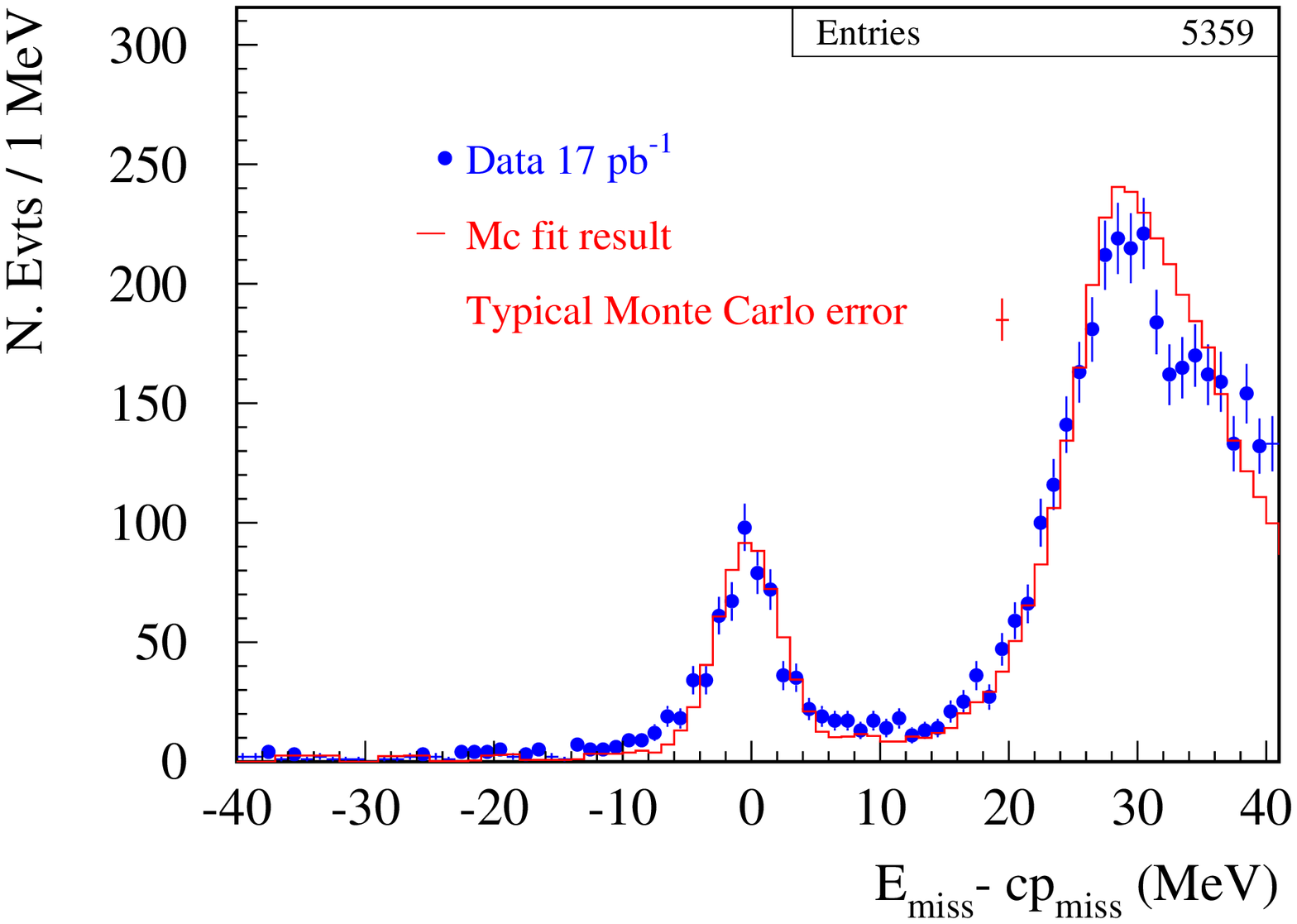, width=0.4\textwidth}
    \caption{$\Emiss-c\Pmiss$ spectrum for \DKSeIII\ candidates before (left) and after (right) TOF cuts.}
    \label{fig:Ke3}
\end{figure}

Finally, for events passing all of the above criteria, we compute the missing energy and momentum \Emiss, \Pmiss. 
For \pienu\ decays, these variables are the neutrino energy and momentum, and satisfy $\Emiss\!=\!c\Pmiss.$
The distribution of $\Emiss-c\Pmiss$ is shown in \Fig{\ref{fig:Ke3}} before (left) and after (right) the 
time-of-flight cuts are imposed. A clear peak at $\Emiss-c\Pmiss\!=\!0$ is evident in the distribution in the
right panel and corresponds to a clean signal for \DKSeIII.
Events with $\Emiss-c\Pmiss>10$\MeV\ are mostly due to \DKSpippim\ decays 
in which a pion decays to a muon before reaching the tracking volume.
The solid line in the right graph is a fit of the data to the sum of the signal 
and background spectra simulated using the Monte Carlo (MC). 
The free fit parameters are the signal and background normalizations. 
We find:
$$N(\pienu)=624\pm30$$
The quoted error includes the contributions from fluctuations in the signal statistics ($\pm25$), from the
background subtraction ($\pm10$), and from the finite statistics of the MC spectra ($\pm13$).

For both \DKSpippim\ (normalization) and \DKSeIII\ (signal) events, contributions to the tagging and selection 
inefficiencies due to purely geometrical effects have been estimated using MC simulation, while data have been 
used to estimate the corrections for tracking and trigger inefficiencies. For \DKSeIII\ events, the corrections 
for vertex reconstruction and time-of-flight $\Ppi$-$\Pe$ identification inefficiencies have also been evaluated 
using data.

For \DKSpippim\ events, the MC is used to compute the efficiency after application of the fiducial cuts,
since these are fundamentally geometrical.
The single-track reconstruction efficiency is obtained in bins of $(p_{\perp},\theta)$ from subsamples of 
\DKSpippim\ events. The ratio of data and MC reconstruction efficiencies is found to be constant throughout the 
acceptance, and the overall MC efficiency is scaled accordingly. The \ks-\kl\ final state allows
the trigger efficiency to be evaluated directly from data. Because of its high energy, the \kl-crash 
always fires at least one calorimeter sector. Approximately 40\% of the time it fires \textit{two}
sectors, giving rise to a valid trigger by itself. 
Events recognized to have triggered in this way are used to measure the probability that at least 
one \ks\ secondary satisfies the trigger together with the \kl-crash cluster, which defines the 
trigger efficiency in the remaining 60\% of events.
The \DKSpippim\ selection efficiencies are listed in 
table~\ref{effistab}; for more details see ref.~\cite{pipinote}.
\begin{table}[ht]
\center
\caption{Efficiencies for \pic\ and $\pi e\nu$ decays.}
\begin{tabular}{l|c}\hline\hline
  \DKSpippim\             & Efficiency \\ \hline
Fiducial cuts             & $0.5755\pm0.0015$    \\
$T_0$ and trigger         & $0.976\pm0.003$ \\ \hline
  \DKSeIII\        & Efficiency \\ \hline
Fiducial cuts             & $0.297\pm0.005$ \\
$T_0$ and trigger         & $0.922\pm0.004$ \\ 
Track-cluster association & $0.925\pm0.006$ \\
Time of flight            & $0.820\pm0.007$ \\ \hline
Tag efficiency ratio  ($R_{\mathrm{tag}}$)~~~ & $1.024\pm0.007$\\
\hline\hline 
 \end{tabular}                                      
 \label{effistab}
\end{table}

For \DKSeIII\ events, the MC is used to evaluate the contributions to the selection inefficiency from 
the vertex reconstruction, the fiducial cuts, and the $M_{\mathrm{\pi\pi}}$ cut. 
The vertex reconstruction efficiency is also estimated from data, 
using events in which the \kl\ decays into \pienu\ near 
the IP and the \ks\ decays into 2$\pi^0$'s ($\pi^0$'s do not affect tracking in the drift chamber).
The ratio of data and MC efficiencies is used as a scale correction to the MC efficiency.
The tracking efficiencies for MC and data have been measured using \DKSpippim\ events, as stated above. 
The need to extrapolate over a larger interval in the track momenta for \DKSeIII\ events
(especially for the electrons) introduces a larger error on the tracking efficiency correction in this case.

The track-to-cluster association efficiency $\varepsilon_{\mathrm{tca}}$
includes both the probability that the particle reaching the calorimeter (\Ppip, \Ppim, \Pemp, or \Pmump) deposits 
enough energy to be detected, and the probability that the track and cluster are correctly associated during 
reconstruction.
The trigger efficiency $\varepsilon_{\mathrm{trg}}$
is given by the probability that one of the \ks\ clusters together with the \kl-crash
cluster triggers the event. 
The overall corrections for $\varepsilon_{\mathrm{tca}}$, $\varepsilon_{\mathrm{trg}}$, and the 
time-of-flight identification efficiency are obtained using 
a high-purity ($>99.7\%$) sample of \DKLeIII, \DKSpippim\ events in which the \kl\ decays near the IP and the 
\ks\ decay generates the trigger.
Such $\kl\ell3$\textit{-gold} events can be selected using track kinematics only, with essentially no recourse 
to calorimeter information.
Independent estimates for $\varepsilon_{\mathrm{tca}}$ and $\varepsilon_{\mathrm{trg}}$ parameterized in the 
variables of each track can be obtained from samples of \Dphipippimpin, \DKSpippim, and $\kl\ell3$-gold events.
For all of these event types, reconstruction is possible whether or not one of the tracks leaves a cluster in the
calorimeter. 
The efficiencies are then parameterized separately for each particle type, and used to weight the MC-generated 
semileptonic events to obtain an average correction. 
The results obtained with the two methods agree within errors. The overall efficiency 
$\epsilon_{\mathrm{tot}}^{\mathrm{\Ppi\Pe\nu}}$ for the detection of 
signal events is \VS{0.208}{0.004}. The various contributions are separately listed in table~\ref{effistab};
for more details see ref.~\cite{pennote}.

In principle, the \kl-crash identification efficiency cancels out in the ratio of the number of selected 
\DKSeIII\ and \DKSpippim\ events.
In practice, since the event \tzero\ is obtained from the \ks\ and the \kl\ is recognized by its time of flight,
there is a small dependence of the \kl-crash identification efficiency on the \ks\ decay mode.
A correction for this effect is obtained by studying the accuracy of the \tzero\ determination in each case.
For the case of \DKSeIII, this study is performed using $\kl\ell3$-gold events in which the topology of the \DKLeIII\ 
decay is manipulated to reflect the \ks\ decay length distribution, and in which the true \tzero\
is unambiguous (both the \Ppip\ and \Ppim\ from the \ks\ give the same \tzero).
For the case of \DKSpippim, this study is performed using \DKLpippimpin, \DKSpippim\ events, in which the \kl\ 
decay gives an unambiguous estimate of the event \tzero.
Such events can be recognized without knowledge of the event \tzero\ {\em a priori.}
The ratio $R_{\mathrm{tag}}$ of the tagging efficiencies for \DKSeIII\ and \DKSpippim\ is found to differ
from unity by $\ab\!2\%$ (see table~\ref{effistab}).

The value for \BR{\DKSeIII} is obtained by normalizing the number of signal events to the number of \DKSpippim\ events
in the same data set, correcting for the total selection efficiencies and the ratio $R_{\mathrm{tag}}$
of tagging efficiencies, and using the present experimental value for \BR{\DKSpippim}~\cite{Groom:2000in}:
$$
\BR{\DKSeIII}=\frac{N^{\mathrm{\Ppi\Pe\nu}}}{N^{\mathrm{\Ppi\Ppi}}}\times
\frac{\varepsilon_{\mathrm{tot}}^{\mathrm{\Ppi\Ppi}}}
{\varepsilon_{\mathrm{tot}}^{\mathrm{\Ppi\Pe\nu}}}\times\frac{1}{R_{\mathrm{tag}}}\times\BR{\DKSpippim}.
$$
We obtain
$$\BR{\ks\toP\pienu}=\SN{(\VS{\VS{6.91}{0.34_{\rm stat}}}{0.15_{\rm syst}})}{-4},$$
in agreement with the expectation from the Standard Model.
The contributions to the error in our measurement are given in table~\ref{errofine}. 
\begin{table}[ht]
\caption{Fractional errors on \BR{\DKSeIII}.}  
\begin{tabular}{l|c}\hline\hline
 Source                 & Error, \%  \\ \hline
 Statistics, signal and background~~~~~~~~~~~~~~~ & 4.9 \\ \hline
 \multicolumn{2}{c}{\DKSeIII\ selection}             \\ \hline
 Tracking and vertex efficiency             & 1.4    \\
 MC preselection efficiency                 & 0.5    \\ 
 Track-cluster, \tzero\ and trigger         & 0.7    \\ 
 Time-of-flight cuts                        & 0.9    \\ \hline
 \multicolumn{2}{c}{\DKSpippim\ selection}           \\ \hline
 Acceptance                                 & 0.2    \\ 
 Tracking                                   & 0.2    \\ 
 T0 and Trigger                             & 0.3    \\ \hline
Tag efficiency ratio  ($R_{\mathrm{tag}}$)  & 0.7    \\ \hline
 \BR{\DKSpippim}                            & 0.4    \\ \hline
 Total error                                & 5.3    \\ \hline\hline
  \end{tabular}
  \label{errofine}
\end{table}

This analysis demonstrates that KLOE can isolate a very pure sample of
\DKSeIII\ decays. The overall selection efficiency for \DKSeIII\ decays among 
\ks-tagged events is \VS{0.208}{0.004}, due in large part to 
the tight fiducial cuts imposed on the decays of interest.

We thank the DA$\Phi$NE team for their efforts in maintaining low
background running conditions and their collaboration during all
data-taking. We also thank F. Fortugno for his efforts in ensuring
good operations of the KLOE computing facilities. This work was
supported in part by EURODAPHNE, contract FMRX-CT98-0169; 
by the German Federal Ministry of Education and Research (BMBF) contract 06-KA-957; 
by Graduiertenkolleg 'H.E. Phys. and Part. Astrophys.' of 
Deutsche Forschungsgemeinschaft, Contract No. GK 742;
by INTAS, contracts 96-624, 99-37; and by TARI, contract HPRI-CT-1999-00088.
\bibliographystyle{elsart-num}
\bibliography{kstopienu4}
\end{document}